\documentclass[12pt,preprint]{aastex}

\def\etal{et al.\ }

\def\unit #1{\,{\rm #1}}

\def\hubble#1{H_0={#1}\unit{km\,sec^{-1}\,Mpc^{-1}}}

\def\mbh{M_{\rm BH}}
\def\re{r_{\rm e}}
\def\bba{\left<}
\def\eba{\right>}
\def\mean #1{\bba #1 \eba}
\def\mubre{\mu_{\rm b}(<\re)}
\def\mmubre{\mean{\mubre}}
\def\kms{\unit{km\,s^{-1}}}
\def\msol{M_{\odot}}
\def\onlyten#1{10^{#1}}
\def\gne #1#2{\ \vphantom{S}^{\raise-0.5pt\hbox{$\scriptstyle#1$}}_
{\raise0.5pt \hbox{$\scriptstyle#2$}}}

\def\lneq{\gne < \sim}
\def\dequn#1#2{Equations~{\ref{eq:#1}}~and~{\ref{eq:#2}}}
\def\equn #1{Equation~\ref{eq:#1}}
\def\labequn #1{\label{eq:#1}}
\def\const{{\rm constant}}
\shorttitle{SuperMassive Black Hole Fundamental Plane}

\shortauthors{Barway \& Kembhavi}

\begin{document}

\title{
A SuperMassive Black Hole Fundamental Plane for Ellipticals}

\author{
Sudhanshu Barway\altaffilmark{1}
 and Ajit Kembhavi\altaffilmark{2}
}
                                                                                
\affil{Inter University center for Astronomy and Astrophysics,
Post Bag 4, Ganeshkhind\\ Pune 411 007, India}

\altaffiltext{1}{email: sudhan@iucaa.ernet.in}            
\altaffiltext{2}{email: akk@iucaa.ernet.in}

\begin{abstract}
We obtain the coefficients of a new fundamental plane for supermassive
black holes at the  centers of elliptical galaxies, involving measured
central black  hole mass and  photometric parameters which  define the
light distribution.  The galaxies  are tightly distributed around this
mass  fundamental plane,  with improvement  in the  rms  residual over
those obtained  from the  $\mbh-\sigma$ and $\mbh-L$  relations.  This
implies  a strong  multidimensional link  between the  central massive
black hole  formation and global photometric  properties of elliptical
galaxies and  provides an  improved estimate of  black hole  mass from
galaxy data.
\end{abstract}


\section{Introduction}

The existence of  massive black holes (hereafter BH)  at the center of
nearby inactive galaxies, as well  as in the nuclei of active galaxies
and  in quasars,  is  well established.   Observations  based on  high
resolution  data and  reverberation  mapping are  now available  which
allow  measurement of  the  masses of  BH  using different  techniques
(Ferrarese  \&   Ford  2005;   Metzroth  \etal  2006;   Shapiro  \etal
2006). Kormendy \&  Richstone (1995) showed that the  measured BH mass
$\mbh$  is correlated  with the  bulge luminosity  $L$ and  bulge mass
M$_{bulge}$  with rms scatter  $\sim0.5$ dex  in $\log\mbh$  (see also
Magorrian  \etal 1998).  A  tight correlation  between $\mbh$  and the
central velocity  dispersion $\sigma$ of the host  galaxy with smaller
rms scatter of $\sim0.34$ dex  in $\log\mbh$ was reported by Ferrarese
\& Merritt  (2000) and Gebhardt  \etal (2000); however,  the published
estimates  of slope  in $\mbh$-  $\sigma$ relation  span a  wide range
(3.75-5.30,  see  Tremaine \etal  2002).   The  small  scatter of  the
$\mbh$- $\sigma$ relation suggests  that the bulge dynamics (or mass),
rather than the luminosity, is responsible for the tight correlation.

It is believed that massive black  holes play an important role in the
formation  and evolution of  galaxies, and  the growth  of the  BH and
bulges must be linked to  the same physical processes; this results in
BH masses that are related to the properties of host galaxies (Silk \&
Rees 1998;  Haehnelt \& Kauffmann  2000; Adams \etal 2001;  Merritt \&
Poon 2004;  Sazonov \etal 2005).   Graham \etal (2001) and  Marconi \&
Hunt (2003)  have shown that  when bulge parameters are  measured with
sufficient accuracy  using the technique  of bulge-disk decomposition,
the resulting scatter in the $\mbh$-$L$ relation is comparable to that
in  the $\mbh$-$\sigma$ relation  (see also  Graham 2007).  Marconi \&
Hunt  (2003) also  suggested  that a  combination  $\sigma$ and  bulge
effective  radius $\re$  should  be used  to  derive the  correlations
between  $\mbh$  and other  bulge  properties.  Recently, Lauer  \etal
(2006) have  suggested  that the  bulge  luminosity  may  be a  better
indicator of  BH mass than the  bulge velocity dispersion  at the high
mass end  for brightest  cluster galaxies.  However,  in spite  of all
these attempts, our understanding of how the photometric properties of
galaxies and their central BHs  are linked in the process of formation
of galaxies remains unclear.

In  this Letter, we  show that  $\log\mbh$, $\log\re$,  and $\mmubre$,
which is the mean bulge  surface brightness in magnitude within $\re$,
are tightly correlated for  nearby elliptical galaxies having measured
central  BH  masses.   The  scatter  around  the  best  fit  plane  is
significantly  less  than   the  scatter  in  various  two-dimensional
relations. It is also less than  the scatter obtained if BH masses are
estimated  from  the  photometric  parameters of  galaxies  using  the
standard  fundamental plane  for ellipticals  and  the $\mbh$-$\sigma$
relation.  In $\S$2  we provide details about the  samples of galaxies
used in the analysis. We present the results in $\S$3, a discussion in
$\S$4 and in $\S$5 a summary  of the work.  Throughout this Letter, we
use $\hubble{70}$, and express  $\re$ in kiloparsec, $\sigma$ in units
of $\kms$, and mass and luminosity in Solar units.

\section{The Data}
To obtain the photometric scaling relation we have considered a sample
of  20 galaxies  classified as  elliptical  in the  Ferrarese \&  Ford
(2005) galaxy list  with measured  black hole masses.   In Table  1 we
report the relevant data for  this sample. To compare the estimates of
central black  hole masses obtained  from our planar relation  and the
$\mbh$- $\sigma$ and $\mbh$-$L$ relations,  we consider a sample of 22
elliptical galaxies  from the Coma cluster.  This  sample was observed
by Jorgensen  \etal (1992) in the  Johnson $B$ band;  a description of
the data and the global parameters obtained from the images can be had
from the reference.

\section{A New Fundamental Plane for Nearby Ellipticals}

The  $\mbh$- $\sigma$  and $\mbh$-  $L$  relations offer  two ways  to
estimate  the BH  mass from  other  galaxy properties,  and have  been
applied to AGN  (McLure \& Dunlop 2002), BL  Lac objects (Falomo \etal
2002), low-redshift radio galaxies  (Bettoni \etal 2003) and to bright
cluster galaxies  (Lauer \etal 2006; Batcherdor \etal  2006).  We have
revisited the  $\mbh$- $\sigma$ relation  and $\mbh$- $L$  relation by
applying a  bisector linear regression fit (Akritas  \& Bershady 1996)
to  the data  given in  Table 1  for the  sample of  nearby elliptical
galaxies with measured BH masses. The two best fit relations are:
\begin{equation}
\log \mbh = (4.53\pm0.49) \log \sigma - (2.24\pm1.17) 
\labequn{smbh1}
\end{equation}
\begin{equation}
\log \mbh = - (0.56\pm0.06) L_B - (3.10\pm1.51)  
\labequn{smbh2}
\end{equation}
The rms scatter  around the best fit lines above is  0.34 dex and 0.42
dex respectively,  along the $\log\mbh$ axis.  Both  the relations are
in good  agreement with  those in Bettoni  \etal (2003)  and reference
therein, but  the relations are  different from those of  Ferrarese \&
Merritt (2000) and Gebhardt \etal (2000),  as we have used a sample of
nearby ellipticals only. It is  possible that some of the scatter seen
in  $\mbh-\sigma$ relation  and  $\mbh-L$ relation  is  caused by  the
effect  of  a  third  parameter.   This is  supported  by  the  strong
correlation  that we  find between  $\log\mbh$ and  $\log\re$,  with a
correlation  coefficient   $r=0.89$,  which  is   significant  at  the
$99.99$ \%  confidence level  for  19 objects;  Marconi and  Hunt (2003) 
have obtained a similar result.

\begin{figure}
\plottwo{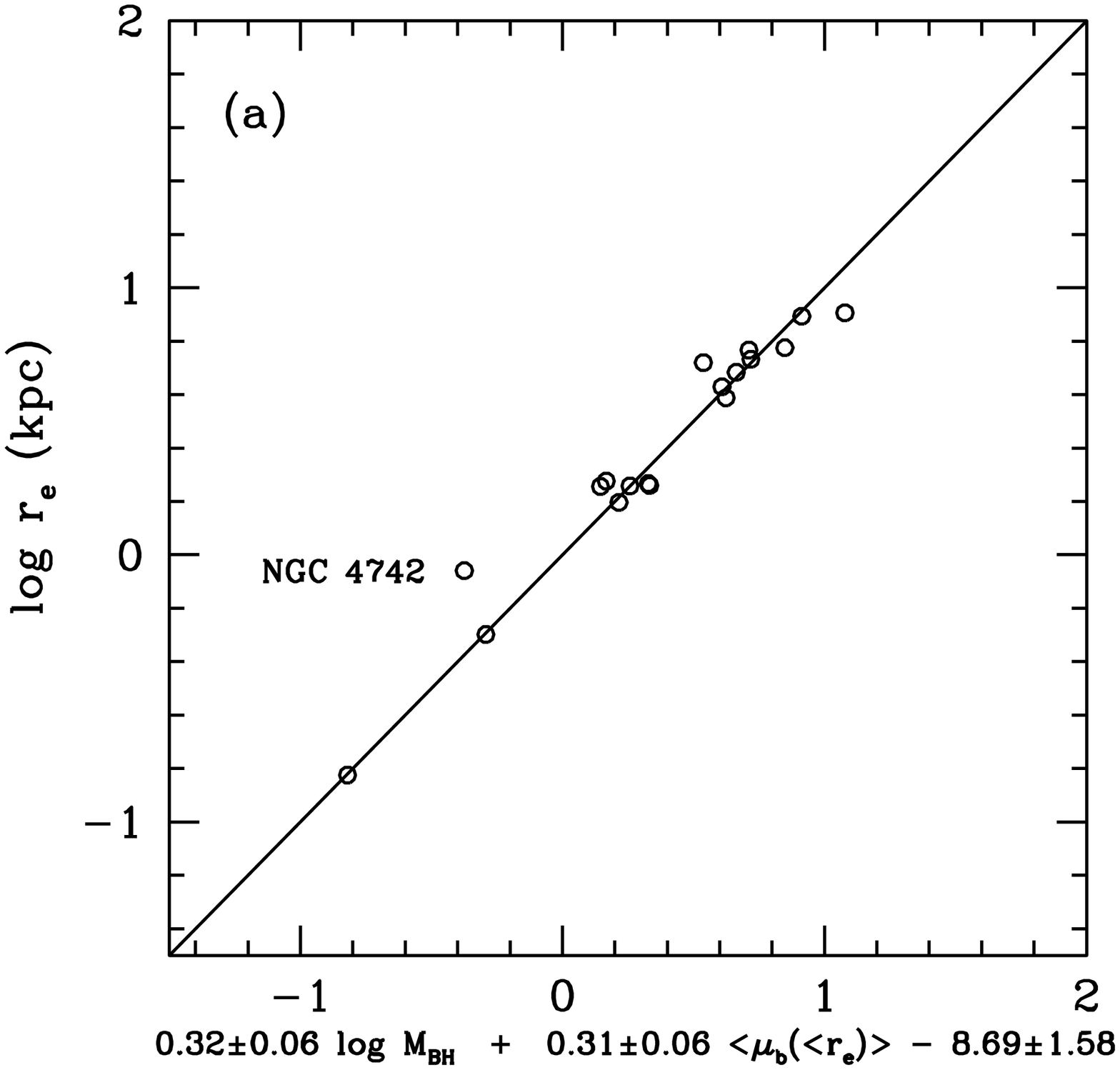}{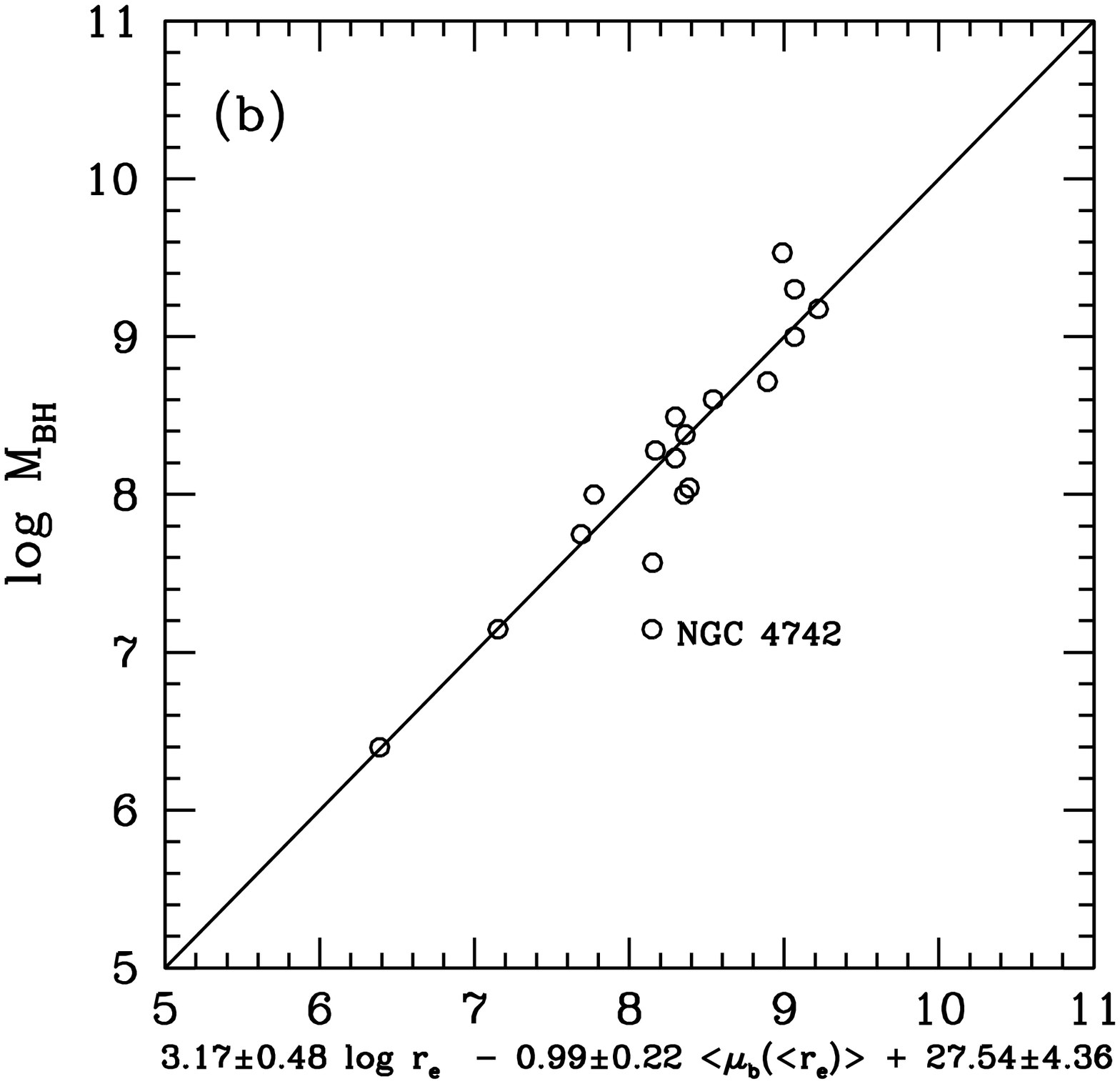}
\caption{Edge-on  views of  the mass  fundamental plane  relations for
nearby ellipticals:  (a) along one of  the shorter axes  of the plane,
$\log\re$ and (b) along another axis of the plane, $\log\mbh$.}
\label{f1}
\end{figure}

Our aim is  to derive a planar relation involving the  BH mass and the
basic photometric parameters $\re$ and  $\mmubre$; this can be used to
estimate the  black hole mass when  it is not  known from measurement,
without reference  to a  spectroscopically measured quantity  like the
central velocity dispersion. We find that the least scatter around the
best-fit plane  in the  space of the  three parameters is  obtained by
expressing it in  the form $\log\re = a\log\mbh  + b\mmubre + \const$.
We minimize  the sum  of the absolute  residuals perpendicular  to the
plane, excluding  one galaxy  NGC\,4742, which is  an outlier  we have
identified  in  Figure~\ref{f1}. The  equation  of  the best-fit  mass
fundamental plane is
\begin{eqnarray}
\log\re &=& (0.32 \pm 0.06) \log\mbh   + (0.31 \pm 0.06) \mmubre \nonumber \\ && \mbox{}
- 8.69 \pm 1.58
\labequn{bhfp}
\end{eqnarray}
%
The uncertainties on the mass  FP coefficients were determined using a
bootstrap  method. An edge-on  view along  log $\re$  of the  plane is
shown  in Figure~\ref{f1}(a).   The rms  scatter in  the  direction of
$\log\re$ is 0.061 dex.  Figure~\ref{f1}(b) shows another edge-on view
of mass  FP in the direction  of $\log\mbh$, with rms  scatter in that
direction of 0.19 dex, which is significantly less than the scatter in
the  $\mbh$-$\sigma$  relation  (Gebhardt  \etal  2000;  Ferrarese  \&
Merritt   2000).    The   outlier   NGC\,4742   is   $6.32\times{({\rm
rms~scatter})}$ from the plane along the $\log\re$ axis.  We have also
obtained the  equation of the  best fit plane including  this outlier.
The rms scatter then increases to  0.078 dex in log $\re$ and 0.25 dex
along log $\mbh$ axis  respectively.  Therefore, even with the outlier
included we  have less scatter  than in the  $\log\mbh-\log\sigma$ and
$\log\mbh-\log L$ fits.

If  we exclude  from the  fit the  four galaxies  NGC\,821, NGC\,2778,
NGC\,4649 and  NGC\,7052 for which the  BH sphere of  influence is not
resolved, and the outlier from  the fit, the rms scatter in $\log\mbh$
around  the best-fit plane  obtained using  the remaining  14 galaxies
decreases to 0.17 dex.

For nearby ellipticals we  have derived the standard fundamental plane
relation,  using  the  same technique  as  in  the  case of  the  mass
fundamental  plane  and  again   excluding  the  outlying  data  point
NGC\,4742. The equation of the best-fit FP is
\begin{eqnarray}
\log\re &=& (1.34 \pm 0.22) \log \sigma  + (0.30 \pm 0.05) \mmubre \nonumber \\ && \mbox{}
- 8.93 \pm 0.74
\labequn{fp}
\end{eqnarray}
The rms scatter is 0.068 dex in log $\re$. The FP coefficients and rms
scatter around  the fit are in  agreement with those  available in the
literature (Jorgensen \etal 2006).
\section{Discussion}
As  suggested by  Ferrarrese  \& Ford  (2005),  given the  photometric
parameters of  an elliptical  galaxy, the central  velocity dispersion
$\sigma$ can be  derived using the FP relation  given in \equn{fp}, if
it is not directly observed, and then the $\mbh$- $\sigma$ relation in
\equn{smbh1} can be  used to estimate the BH  mass. However, the error
in the  estimated BH mass will  then be the cumulative  error of these
two relations,  thus increasing the uncertainty in  the mass estimate.
Another  disadvantage  of this  approach  is  that  the slope  in  the
$\mbh-\sigma$ relation  spans the  range 3.75-5.3, leading  to further
uncertainty  in the  estimate of  the mass.  The mass  FP  provides an
improvement over this two step  procedure, and also helps to constrain
the slope of the $\mbh$- $\sigma$ relation, as described below.

\begin{figure}
\plotone{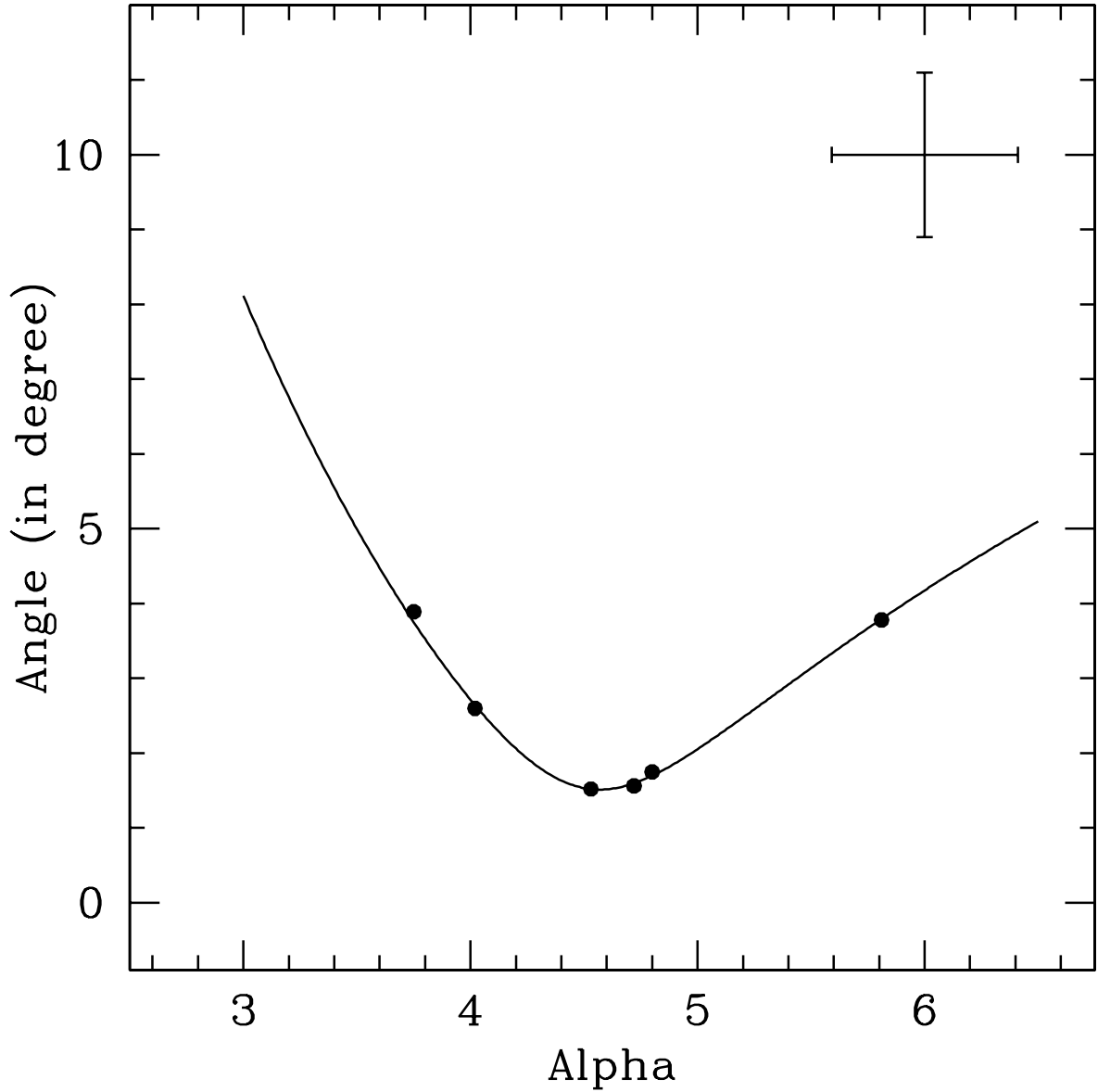}
\caption{The curve shows  the angle between (1) the  best-fit mass FP,
and (2) the plane derived using the fundamental plane in \equn{fp} and
the relation $\mbh= \alpha\log\sigma + \beta$ for a range of values of
$\alpha$.  The filled circles indicate  the angle for actual values of
$\alpha$  taken from  the literature  (see Tremain  \etal  2002).  The
typical error in the measured values of $\alpha$ and the derived angle
between the planes is shown at the top right in the plot.}
\label{f2}
\end{figure}

We   consider   a   two-dimensional   relation   of   the   form   log
$\mbh=\alpha\log\sigma  +\beta$,   where  $\alpha$  and   $\beta$  are
constants to  be determined.  Introducing this into  \equn{fp} for the
fundamental  plane,  we  get  a  plane in  the  space  of  $\log\mbh$,
$\log\re$ and $\mmubre$, with the direction of the normal to the plane
dependent  on  the value  of  $\alpha$.   In  Figure~\ref{f2} we  have
plotted,  as a  solid line,  the angle  between this  normal,  and the
normal  to the  mass  FP in  \equn{bhfp},  for a  range  of values  of
$\alpha$.   The  filled  circles  on  the curve  indicate  the  angles
corresponding to specific values  of $\alpha$ found in the literature,
obtained by various  groups from their fits to  the data (see Tremaine
\etal 2002).   It is seen from  the figure that the  angle between the
two planes is minimum near  $\alpha=4.5$, which should be the value to
be used in the  $\log\mbh-\log\sigma$ relation to determine black hole
mass  from   the  central  dispersion  velocity.   The   best  fit  in
\equn{smbh1} corresponds to $\alpha=4.53$.   It will be interesting to
see how  the minimum  value of $\alpha$  depends on  the morphological
type of the host galaxy.

\begin{figure}
\plotone{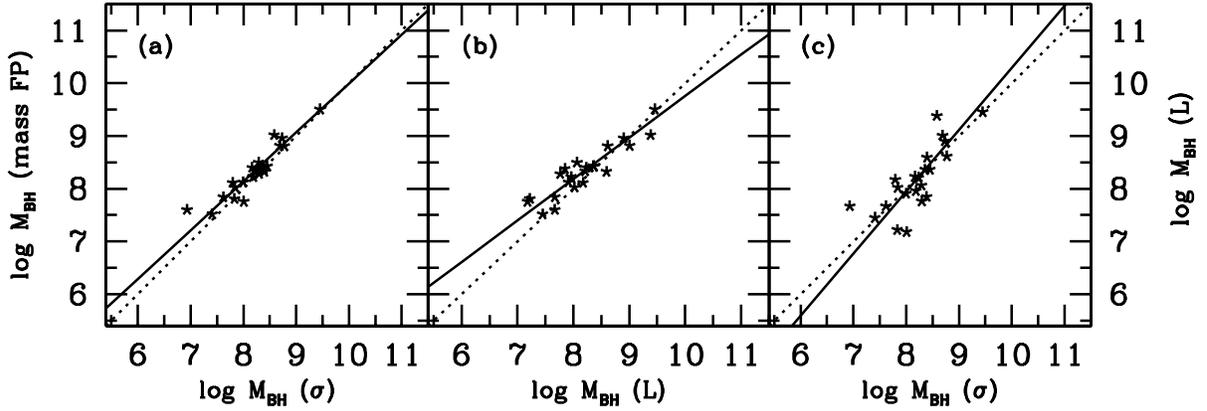}
\caption{Comparison  of  black  hole  mass estimated  using  the  mass
fundamental plane, $\mbh(mass \ FP)$,  with (a) the mass $\mbh(\sigma)$
estimated using \equn{smbh1} and $\mbh(\sigma)$ and (b) with the mass
$\mbh(L)$  estimated  using \equn{smbh2}.   In  panel  (c) we  compare
$\mbh(\sigma)$ with $\mbh(L)$.  In  each panel the dark line indicates
the linear  fit to the points  shown, while the dashed  line has slope
unity.}
\label{f3}
\end{figure}

We have used the  mass FP to predict the black hole  mass for a set of
22 elliptical  galaxies from the Coma cluster,  using photometric data
from Jorgensen  \etal (1992).   We have also  obtained the  black hole
mass  for  these   galaxies  using  \dequn{smbh1}{smbh2}.  The  masses
obtained in these various ways  are compared in Figure~\ref{f3}. It is
seen  from Figure~\ref{f3}(a)  that the  agreement  between $\mbh({\rm
mass~FP})$  and $\mbh(\sigma)$  is  good; the  points are  distributed
around  a  line  with  slope   close  to  unity,  with  a  correlation
coefficient  $r=0.93$,  which  is   significant  at  better  than  the
$99.9$ \%  level.  A larger number  of points will be needed for a
better comparison  and to examine any departures  from linearity.  The
slope in  the $\log\mbh-\log\sigma$ relation in  \equn{smbh1} is close
to the minimum value  of $\alpha$ obtained from Figure~\ref{f2}. Using
any other value of $\alpha$  will produce a less favorable comparison.
We see  from panel  (b) that $\mbh({\rm  mass~FP})$ and  $\mbh(L)$ are
distributed  along a  straight line  with slope  less than  unity; for
$\mbh\lneq\onlyten{8.5}\msol$,  masses obtained from  the FP  would be
systematically less  than masses  obtained from the  $\log\mbh-\log L$
relation.  The  dispersion of the points  around the best  fit line is
greater  in this case  than in  panel (a). We have  for completeness
compared   in  panel  (c)   black  hole   masses  obtained   from  the
$\log\mbh-\log\sigma$  and $\log\mbh-\log  L$  relations respectively,
and find a slope greater than  unity and larger dispersion than in the
other cases.

The three  dimensional mass  FP has lesser  rms deviation than  in the
earlier two dimensional relations while some reduction in residuals is
expected when  the number of parameters  in the fit  is increased from
two  to three,  it  appears that  the  process can  not  be taken  any
further.  We  have  considered  a  four  dimensional  plane  with  the
dispersion velocity  $\sigma$ included in  the fit along with  the two
photometric parameters.  However, we  find that the residuals from the
three dimensional plane are  not correlated with $\log\sigma$, and the
quality  of a  four dimensional  fit involving  $\log\mbh$, $\log\re$,
$\mmubre$ and  $\log\sigma$ is poor.  A three  dimensional relation is
therefore the best we can do with the available data.

It will be  interesting to obtain the mass FP  for photometric data in
the near-infrared bands, since stellar population metallicity effects
are  less important  than  in  the optical  region (Pahre \etal 1998).  
Another issue  to
examine  is whether the  bulges of  galaxies of  various morphological
types share a common mass FP.

\section{Summary}
We  have shown  that $\log\re$,  $\log\mbh$ and  $\mmubre$  for nearby
elliptical  galaxies having  measured  central BH  masses are  tightly
distributed  about  a plane  with  a rms  scatter  of  0.19 dex  along
$\log\mbh$.  The scatter  decreases to 0.17 dex in  $\log\mbh$ when we
use  only those  galaxies  for which  the  BH sphere  of influence  is
resolved. The mass FP provides a convenient way for estimating BH 
mass from photometric  data alone.

\acknowledgments 
We  thank   Swara  Ravindranath  and  Yogesh   Wadadekar  for  helpful
discussions. We  also thank the  referee, Luca Ciotti,  for insightful
comments,  which  helped to  improve  the  original manuscript.   This
research has  made use of the NASA/IPAC  Extragalactic Database (NED),
which  is  operated  by  the  Jet  Propulsion  Laboratory,  California
Institute of Technology, under  contract with the National Aeronautics
and Space Administration.

\clearpage

\begin{deluxetable}{llrccrrc}
\tabletypesize{\small}
\tablewidth{0pt} 
\tablecaption{Basic parameters for elliptical galaxies with measured black hole mass.}
\tablecolumns{8}
\tablehead{
\colhead{Object} &
\colhead{Type} &
\colhead{Distance} &
\colhead{$\mbh$} &
\colhead{$\sigma$} &
\colhead{$L_B$} &
\colhead{log r$_e$} &
\colhead{$\mmubre$} 
  \\
\colhead{} &
\colhead{} &
\colhead{(Mpc)} &
\colhead{(10$^8$ M$_{\odot}$)} &
\colhead{(km s$^{-1}$)} &
\colhead{(mag)} &
\colhead{(kpc)} &
\colhead{(mag arcsec$^{-2}$)} 
}
\startdata
NGC 221/M32   &  $-$6.0  & 0.80  & $2.5^{+0.5}_{-0.5}$ $\times$ 10$^6$ &\ 75$\pm$10 & $-$15.80$\pm$0.18 & $-$0.83  & 18.69 \\
NGC 821       &	 $-$5.0  & 24.1  & $3.7^{+2.4}_{-0.8}$ $\times$ 10$^7$ & 209$\pm$26 & $-$20.42$\pm$0.21 &  0.72    & 21.85 \\
NGC 2778      &	 $-$5.0  & 22.9  & $1.4^{+0.8}_{-0.9}$ $\times$ 10$^7$ & 175$\pm$22 & $-$18.58$\pm$0.33 &  0.26    & 21.38 \\
NGC 3377      &	 $-$5.0  & 11.2  & $1.0^{+0.9}_{-0.1}$ $\times$ 10$^8$ & 145$\pm$17 & $-$19.18$\pm$0.13 &  0.26    & 20.76 \\
NGC 3379      &	 $-$5.0  & 10.6  & $1.0^{+0.6}_{-0.5}$ $\times$ 10$^8$ & 206$\pm$26 & $-$19.81$\pm$0.20 &  0.26    & 20.16 \\
NGC 3608      &	 $-$5.0  & 22.9  & $1.9^{+1.0}_{-0.6}$ $\times$ 10$^8$ & 182$\pm$27 & $-$20.07$\pm$0.17 &  0.59    & 21.41 \\
NGC 4261      &	 $-$5.0  & 31.6  & $5.2^{+1.0}_{-1.1}$ $\times$ 10$^8$ & 315$\pm$38 & $-$21.23$\pm$0.20 &  0.77    & 21.25 \\
NGC 4291      &	 $-$5.0  & 26.2  & $3.1^{+0.8}_{-2.3}$ $\times$ 10$^8$ & 242$\pm$35 & $-$19.72$\pm$0.35 &  0.27    & 20.25 \\
NGC 4374/M84  &	 $-$5.0  & 18.4  & $1.0^{+2.0}_{-0.6}$ $\times$ 10$^9$ & 296$\pm$37 & $-$21.40$\pm$0.31 &  0.68    & 20.81 \\
NGC 4473      &	 $-$5.0  & 15.7  & $1.1^{+0.4}_{-0.8}$ $\times$ 10$^8$ & 190$\pm$25 & $-$19.86$\pm$0.14 &  0.28    & 20.19 \\
NGC 4486/M87  &	 $-$4.0  & 16.1  & $3.4^{+1.0}_{-1.0}$ $\times$ 10$^9$ & 375$\pm$45 & $-$21.71$\pm$0.16 &  0.91    & 21.60 \\ 
NGC 4564      &	 $-$5.0  & 15.0  & $5.6^{+0.3}_{-0.8}$ $\times$ 10$^7$ & 162$\pm$20 & $-$18.94$\pm$0.18 &  0.19    & 20.64 \\
NGC 4697      &	 $-$5.0  & 11.7  & $1.7^{+0.2}_{-0.1}$ $\times$ 10$^8$ & 177$\pm$10 & $-$20.20$\pm$0.18 &  0.63    & 21.41 \\
NGC 4649/M60  &	 $-$5.0  & 16.8  & $2.0^{+0.4}_{-0.6}$ $\times$ 10$^9$ & 385$\pm$43 & $-$21.30$\pm$0.16 &  0.78    & 21.10 \\
NGC 4742      &	 $-$5.0  & 15.5  & $1.4^{+0.4}_{-0.5}$ $\times$ 10$^7$ &\ 90$\pm$05 & $-$19.03$\pm$0.10 & $-$0.06  & 19.36 \\
NGC 5845      &	 $-$5.0  & 25.9  & $2.4^{+0.4}_{-1.4}$ $\times$ 10$^8$ & 234$\pm$36 & $-$18.92$\pm$0.25 & $-$0.30  & 18.38 \\
NGC 7052      &	 $-$5.0  & 71.4  & $4.0^{+2.8}_{-1.6}$ $\times$ 10$^8$ & 266$\pm$34 & $-$21.43$\pm$0.38 &  0.89    & 22.01 \\
IC 1459	      &	 $-$5.0  & 29.2  & $1.5^{+1.0}_{-1.0}$ $\times$ 10$^9$ & 340$\pm$41 & $-$21.45$\pm$0.32 &  0.73    & 20.81 \\
NGC 6251      &	 $-$5.0  & 107.0 & $6.1^{+2.0}_{-2.1}$ $\times$ 10$^8$ & 290$\pm$39 & $-$21.95$\pm$0.28 &  1.31    &  ---  \\
CygA          &	 $-$5.0  & 240.0 & $2.9^{+0.7}_{-0.7}$ $\times$ 10$^9$ & 270$\pm$90 & $-$20.03$\pm$0.27 &   ---    &  --- 
\enddata
\tablecomments{Cols. 1 and 2 give the name and the morphological 
type from RC3; Col. 3 the distance, derived from Surface Brightness Fluctuations 
(SBF, Tonry \etal 2001); Cols. 4-6 provide the adopted values for the mass of 
black hole $\mbh$, velocity dispersion and absolute bulge luminosity $L$ in $B$ band
(from Ferrarese \& Ford 2005); Cols 7 and 8 give the effective radius $\re$ 
(from Faber \etal 1989 and using the distance in Col. 3) and mean surface brightness
within  effective radius in $B$ band (from Faber \etal 1989).}
\end{deluxetable}

\clearpage

\clearpage

\clearpage

\end{document}